\documentstyle[12pt]{article}
\newcommand{\be}{\begin{equation}}
\newcommand{\en}{\end{equation}}
\newcommand{\bea}{\begin{eqnarray}}
\newcommand{\ena}{\end{eqnarray}}

\newcommand{\hbo}{\hbox to 1 true cm {\hfill } }
\newcommand{\tr}{\hbox{tr}}

\def\dslash{\partial\kern-.5em\slash}
\def\kslash{k\kern-.5em\slash}

\begin{document}
\vglue 1truecm

\vbox{ UNIT\"U-THEP-10$/$1993
\hfill August 19, 1993.
}

\vfil
\centerline{\bf \large Non-trivial SU(N) instantons and exotic skyrmions$^1$}

\bigskip
\centerline{ H.\ Reinhardt, K.\ Langfeld }
\medskip
\centerline{Institut f\"ur theoretische Physik, Universit\"at
   T\"ubingen}
\centerline{D--74076 T\"ubingen, Germany }
\bigskip

\vfil
\begin{abstract}

The classical Yang-Mills equations are solved for arbitrary semi-simple
gauge groups in the Schwinger-Fock gauge.
A new class of SU(N) instantons is presented
which are not embeddings of SU(N-1) instantons but have non-trivial
SU(N) color structure and carry winding number $n=N(N^{2}-1)/6$.
Explicit configurations are given for SU(3) and SU(4) gauge groups.
By means of the Atiyah Manton procedure Skyrmion fields are constructed
from the SU(N) instantons. These Skyrmions represent exotic baryon states.

\end{abstract}

\vfil
\hrule width 5truecm
\vskip .2truecm
\begin{quote}
$^1$ Supported by DFG under contract Re $856/1 \, - \, 1$
\end{quote}
\eject

{\it 1.\ Introduction \/ }

Instantons~\cite{inst} are classical finite action solutions to Euclidean
field theory describing quantum tunneling. They play an important
role in a semiclassical understanding of the ground state of
complex quantum systems within the path integral approach~\cite{ra82}.
In Yang Mills (YM) theories instantons describe the tunneling
between topologically different sectors~\cite{inst}.
In particular, in QCD
instantons are considered to be responsible for the complex
structure of the QCD vacuum and constitute the basis for various
models of the QCD ground state, as e.g.\ the dilute instanton
gas model~\cite{ca78}
or the instanton liquid model~\cite{sh82}. In these approaches
instantons provide a mechanism for spontaneous breaking of chiral
symmetry~\cite{ca78,sh82} leading to the successful interpretation
of pseudoscalar mesons as Goldstone bosons. Furthermore instantons
offer a solution to the U(1) problem~\cite{tho86,tho76b}.
So far all explicit instanton calculations, in particular the
instanton models of the QCD ground state,
are based on the celebrated Polyakov-'t~Hooft instanton of
the SU(2) Yang-Mills theory or more precisely on its  SU($N>2$)
embedding. An important and hitherto open question is whether
in SU($N>2$) gauge theories instantons
which are not SU(N) embeddings of the SU(2) instanton
are important for an understanding of the SU($N$) Yang-Mills vacuum,
in particular of the QCD ground state. In order to answer this question
explicit representations of instantons are required.
There is a general method based on works of Atiyah et
al.\ \cite{at77,co78} which, in principle, allows one to find all (anti-)
self-dual instanton solutions, although in practice it might be
difficult to obtain explicit representations.

In this letter we follow a different route by directly
solving the Yang-Mills equations of motion. We report on novel
SU($N$) instantons which are not
SU($N$) embeddings of subgroup instantons but have
non-trivial SU($N$) group structure. For SU($N$) we will find (anti)
self-dual instantons up to winding number $n =\pm N(N^{2}-1)/6$.
We will give explicit
representations of  these instantons for the gauge groups SU(3) and
SU(4).

Atiyah and Manton have considered SU(2) Skyrmion fields
from the standard SU(2) instanton by evaluating the holomony of the
instanton along the time axis. We generalise their method to the
non-trivial SU(N) instanton configurations and construct exotic Skyrmions
with baryon number larger than one.

\bigskip
{\it 2.\ Instantons in Fock-Schwinger gauge }

\medskip
We are looking for extrema of the classical Euclidean
Yang Mills action
\be
S [F] = \frac{ 1 }{ 4 g^{2} } \int d^{4}x \; F^{a}_{\mu \nu}
F^{a}_{\mu \nu } \; .
\label{eq:1}
\en
Here $g$ is the coupling constant and
\be
F^{a}_{\mu \nu } \; = \; \partial_{\mu } A^{a}_{\nu } \, - \,
\partial_{\nu} A^{a}_{\mu } \, + \, f^{abc} A^{b}_{\mu } A^{c}_{\nu } \; ,
\label{eq:2}
\en
is the field strength of the gauge potential $A^{a}_{\mu }$.
Furthermore $f^{abc}$ denotes the structure constant of the gauge
group $\cal G$.
We will focus on SU($N$) gauge theories, but most of our considerations
given below apply to any semi-simple Lie group.
Non-trivial solutions to classical Yang-Mills-theories are known
to exist only in four dimensions~\cite{deser}, consequently
we stick to $D=4$.
Extremising the action
yields the classical equation of motion
\be
\partial _{\mu } F^{a}_{\mu \nu } \; = \;  \hat{F}^{ab}_{\nu \mu }
A^{b}_{\mu} \; ,
\hbox to 1 true cm {\hfill}
\hat{F}^{ab}_{\mu \nu } = f^{abc} F^{c}_{\mu \nu } \; .
\label{eq:1a}
\en
Any finite action gauge potential $A^{a}_{\mu }(x)$, in particular
the instantons, must become pure gauge $U \partial _{\mu } U^{\dagger }$
as $x^{2}$ tends to infinity. They can be classified
by the so called topological charge
\be
n [F] \; = \; \frac{ 1 }{ 16 \pi ^{2}  } \int d^{4}x \;
\tr ( F_{\mu \nu } F^{\ast}_{\nu \mu } )
\label{eq:6i}
\en
which is  an integer (the Pontryagin index).
The Bianchi-identity
\be
\partial _{\mu } F^{a \, \ast }_{\mu \nu } \; = \; (\hat{F} ^{\ast }
 )^{ab}_{\mu \nu } A^{b}_{\mu } \; , \hbo
F^{a \, \ast }_{\mu \nu } \; = \; \frac{1}{2} \epsilon _{\mu \nu
\alpha \beta } F^{a  }_{\alpha \beta } \; ,
\label{eq:2a}
\en
which holds for arbitrary gauge fields, implies that any gauge field
which has an (anti-) self dual field strength (i.\ e.\
$F^{a \, \ast }_{\alpha \beta } = (-) F^{a }_{\alpha \beta }$ ),
represents an instanton solution to the classical YM equation (\ref{eq:1a}),
in agreement with the standard variational result~\cite{inst}.

For a given gauge potential $A^{a}_{\mu } (x)$ however, we do not know
whether its field strength $F^{a}_{\mu \nu }$ is self-dual
until we actually calculate it. Thus in the search for instantons,
it is much more convenient to consider the field strength
as an independent dynamical variable instead of the gauge potential
itself. This can be accomplished in the Fock-Schwinger gauge
\be
x_{\mu }  A_{\mu }(x)=0 \; .
\label{gauge}
\en
In this gauge,
the gauge potential can be entirely expressed in terms of the
field strength i.\ e.\ ,
\be
A^{a}_{\mu }(x) \; = \; - \int_{0}^{1} d\alpha \;
\alpha \; F^{a}_{\mu \nu }(\alpha x ) \; x_{\nu }  \; .
\label{eq:a1}
\en
and the equation of motion (\ref{eq:1a}) becomes an
equation for $F^{a}_{\mu \nu }$ i.\ e.\ ,
\be
\partial _{\mu } F^{a}_{\mu \nu }(x) \; = \; - \hat{F}^{ab}_{\nu \lambda }
(x) \int_{0}^{1} d \alpha \; \alpha \, F^{b}_{\lambda \sigma }
x_{\sigma } \; .
\label{eq:a1a}
\en
This equation is equivalent to the original equation of motion only
for those $F^{a}_{\mu \nu } $ which are the field strengths to some
gauge potential. For an arbitrary $F'^{a}_{\mu \nu }$
equation (\ref{eq:a1}) yields a gauge potential $A^{a}_{\mu }[F']$
whose field strength $F^{a}_{\mu \nu }\left[ A[F'] \right] $
differs in general from the initial
$F'^{a}_{\mu \nu }$. We must therefore supplement (\ref{eq:a1a})
with the consistency requirement
\be
F^{a}_{\mu \nu } \bigl[ A[F] \bigr] \; = \; F^{a}_{\mu \nu } \; .
\label{eq:constr}
\en
With this constraint, (\ref{eq:a1a})
is completely equivalent to the classical YM equation
of motion (\ref{eq:1a}) that we started with.

We now solve (\ref{eq:a1a}) with the isotropic ansatz
\be
F^{a}_{\mu \nu } \; = \; G^{a}_{\mu \nu } \, \psi (x^{2}) \; ,
\label{eq:a2}
\en
where $G^{a}_{\mu \nu } $ is a constant, (anti-) self-dual matrix
antisymmetric under the exchange of $\mu $ and $\nu $.
For this ansatz the gauge field (\ref{eq:a1}) becomes
\be
A^{a}_{\mu } \; = \; - G^{a}_{\mu \nu } x_{\nu } \frac{1}{2 x^{2} }
\int_{0}^{x^{2} } du \; \psi (u) \; =: \;
- G^{a}_{\mu \nu } x_{\nu } \Phi ( x^{2} ) \; .
\label{eq:a3}
\en
and the equation of motion (\ref{eq:a1a}) reduces to
\be
- 2 G^{a}_{\mu \nu } x_{\nu } \, \psi ' \; + \; \hat{G}^{ab}_{\mu \nu }
G^{b}_{\nu \omega } x_{\omega } \, \Phi (x^{2}) \psi (x^{2}) \; = \; 0 \; .
\label{eq:a5}
\en
This equation makes it easy to separate
the color and Lorentz structure from the space-time
dependence, reducing (\ref{eq:a5})
(uniquely up to unimportant rescaling of $G \rightarrow G/ \beta $
and $\Phi \rightarrow \beta \Phi $ with an arbitrary constant $\beta $) to
\bea
\hat{G}^{ab}_{\mu \lambda } G^{b}_{\lambda \nu } &=& 2 \, G^{a}_{\mu \nu }
\label{eq:a6} \\
\psi '( x^{2} ) & = & \Phi (x^{2}) \, \psi (x^{2}) \; .
\label{eq:a7}
\ena
%...................
%
In order to solve the matrix equation (\ref{eq:a6}) it is convenient to expand
the antisymmetric color matrices $G^{a}_{\mu \nu }$ in terms
of 't~Hooft's $\eta $-symbols i.\ e.\ ,
\be
G^{a}_{\mu \nu } \; = \; G^{a}_{i} \eta^{i}_{\mu \nu } \; + \;
\bar{G} ^{a}_{i} \bar{\eta }^{i}_{\mu \nu } \; .
\label{eq:14i}
\en
The $\eta ^{i} $ and $\bar{\eta }^{i}, \; (i=1,2,3)$ are self-dual
and anti self-dual space-time tensors, which generate the
$SU(2) \times SU(2) \sim SO(4)$ symmetry group of
Euclidean space, satisfying
\be
[ \eta ^{i}, \eta ^{k} ] = -2 \epsilon ^{ikl} \eta ^{l} \; ,
\hbox to 1 true cm {\hfill}
[ \bar{\eta } ^{i}, \bar{\eta } ^{k} ] = -2 \epsilon ^{ikl}
\bar{\eta } ^{l} \; ,
\hbox to 1 true cm {\hfill}
[ \eta ^{i}, \bar{\eta }^{k} ]=0  \; ,
\label{eq:15i}
\en
\be
\{ \eta ^{i}, \eta ^{k} \} \; = \; - 2 \delta ^{ik} \; , \hbo
\{ \bar{\eta }^{i}, \bar{\eta }^{k} \} \; = \; - 2 \delta ^{ik} \; , \hbo
\tr \{ \eta^{i} \bar{\eta }^{k} \} \; = \; 0 \; .
\nonumber
\en
For the moment let us confine ourselves to self-dual configurations
i.\ e.\ , $\bar{G}^{a}_{i}=0$.
Using (\ref{eq:15i}) the relation (\ref{eq:a6}) can be cast in the form
\be
G^{a}_{l} = \frac{1}{2} \epsilon _{lrs} f^{abc} G^{b}_{r} G^{c}_{s} \; .
\label{eq:32}
\en

We must also guarantee that the field strength ansatz (\ref{eq:a2})
is constructed from the gauge potential (\ref{eq:a3}).
So, our task is to solve the constraint (\ref{eq:constr}) together with
the e.o.m.'s (\ref{eq:a7}) and (\ref{eq:32}) for $F^{a}_{\mu \nu }$
and $A^{a}_{\mu }$ given by (\ref{eq:a2}) and (\ref{eq:a3})
respectively.
Decomposing the constraint (\ref{eq:constr}) into self-dual and
anti self-dual parts one finds
\bea
G^{a}_{i} ( 2 \Phi + x^{2} \Phi ' ) \, + \, \frac{1}{4}
\epsilon _{ikl} f^{abc} G^{b}_{k} G^{c}_{l} \, x^{2} \Phi ^{2}  &=&
G^{a}_{i} \psi
\nonumber \\
- x_{\nu } \bar{\eta } ^{i}_{\nu \mu } \eta ^{k}_{\mu \omega }
x_{\omega } \, ( G^{a}_{k} \Phi ' - \frac{1}{4}
\epsilon _{kml} f^{abc} G^{b}_{m} G^{c}_{l} \, \Phi ^{2} ) &=& 0 \; .
\nonumber
\ena
Using the relation (\ref{eq:32}) of the coefficients $G^{a}_{i}$, we
can reduce these to ordinary differential equations i.\ e.\ ,
\bea
x^{2} \Phi ' \, + 2 \Phi \, + \, \frac{1}{2} x^{2} \Phi ^{2}
&=& \psi
\label{eq:a15} \\
\ \Phi ' \, - \, \frac{1}{2} \Phi ^{2} &=& 0 \; .
\label{eq:a16}
\ena
These equations are solved by
\be
\Phi (x^{2}) \; = \; - \frac{2}{ x^{2} + \rho ^{2} } \; , \hbo
\psi (x^{2}) \; = \; - \frac{ 4 \rho ^{2} }{ ( x^{2} + \rho ^{2} )^{2} }
\label{eq:a17}
\en
where $\rho $ is an arbitrary parameter.
It is not surprising that these solutions also satisfy (\ref{eq:a7}),
which follows from the classical YM equation (\ref{eq:1a})
within the gauge (\ref{gauge}) and with the ansatz (\ref{eq:a2}).
This is merely a consequence of the general fact that
any (anti-) self-dual field configuration satisfies the classical
YM equations. In fact, the solutions (\ref{eq:a17}) have precisely the
space-time dependence of the gauge potential and the field strength,
respectively, of the standard Polyakov-t'Hooft instanton.

What remains to be done is to solve the algebraic equation
(\ref{eq:32}) which defines the color and Lorentz structure
of the instantons.
Let $t^{a}$ denote the generators
of the gauge group $\cal G$ satisfying $[t^{a},t^{b}]=if^{abc} t^{c}$ in
the fundamental representation. Defining
\be
G_i \; = \; G^{a}_{i} t^{a}
\label{eq:18i}
\en
the algebraic equation (\ref{eq:32})  can be rewritten as
\be
[ G_{j}, G_{k} ]  \; = \; i \epsilon _{jkl} G_{l} \; ,
\label{eq:34}
\en
Since the $\epsilon ^{ijk}$ are the structure constants of the SU(2)
group, any realisation $G_{i}$
of the SU(2) algebra yields an instanton solution.

For field configurations (\ref{eq:a2}) with radial dependence
(\ref{eq:a17}) the winding number and the corresponding action become
\be
n \; = \; \frac{1}{3} \,  G^{a}_{i} G^{a}_{i} \; = \;
\frac{2}{3} \tr G_{i} G_{i} \; ,
\hbox to 1 true cm {\hfill}
S \; = \; \frac{ 8 \pi ^{2} }{ 3 g^{2} } \,  G^{a}_{i} G^{a}_{i} \; = \;
\frac{4 \pi ^{2} }{g^{2}} n \; .
\label{eq:20i}
\en
Noting that instanton solutions $G_{i}$
are representations of the SU(2) spin algebra for some spin $s$,
the quantity
$G_{i} G_{i}$ represents the quadratic
casimir operator of the SU(2) group, i.\ e.\ ,
\be
G_{i} G_{i} \; = \;  s(s+1) \; 1_{G} \; =: \; c(G) \, 1_{G}
\hbox to 2 true cm {\hfil with \hfil }
\tr \, 1_{G} =: d(G) \; .
\label{eq:21i}
\en
$1_{G}$ is the unit matrix in the spin representation defined
by the $G_{i}$,
which need not to coincide with the $N$-dimensional unit matrix of the color
representation but may have lower dimension.
The winding number and action can therefore be expressed as
\be
n= \frac{2}{3}  c(G) d(G) \; ,
\hbox to 1 true cm {\hfill}
S=\frac{16 \pi^{2} }{3 g^{2} }  c(G) d(G)  \; .
\label{eq:23i}
\en

\bigskip
{\it  3.\ Explicit construction of SU(N) instantons }

\medskip
Let us now construct explicit solutions to (\ref{eq:34}), which
define the color structure of the instantons.
In the fundamental representation of SU($N$) the $t^{a}$
form a complete basis for the hermitean $N \times N$ matrices.
Furthermore, the lowest dimensional irreducible spin $s$ representation
of SU(2) is realised by hermitean $(2s+1) \times (2s+1)$
matrices. Hence, the N-dimensional $s=(N-1)/2$ spin representation always
provides an instanton $G_{i}$ with non-trivial SU($N$) color structure.
We refer to this instanton as the instanton of maximum spin (for given $N$).
{}From (\ref{eq:23i}) it follows that this
instanton carries winding number
\be
n \; = \; \frac{2}{3} \, N \, s(s+1) \; = \;
\frac{2}{3} N \, \frac{N-1}{2} \, \frac{N+1}{2} \; = \;
\frac{1}{6} N(N^{2}-1) \; .
\label{eq:29i}
\en
Additional instantons will usually arise from $N$-dimensional
realisations of lower spin $s<(N-1)/2$ representations. In particular,
all embeddings of SU($N-1$) instantons in SU($N$) obviously represent
SU($N$) instanton configurations.

For the gauge group SU(2) any representation $t^{a}$ of the color
generators naturally provides an instanton by identifying
the color group with the spin group, i.\ e.\ ,
\be
G^{a}_{i} \; = \; \delta ^{a}_{i} \hbo
(\hbox{ identity map }) \; .
\label{eq:24i}
\en
Eq.(\ref{eq:a2},\ref{eq:14i}) then yields
the standard Polyakov-'t~Hooft instanton
$(G^{a}_{\mu \nu }= \eta ^{a}_{\mu \nu })$.
This is the instanton with maximal spin for SU(2).
Obviously, in this case, there are
(up to global color and Lorentz transformations)
no mappings of the SU(2) color
group into the SU(2) spin algebra other than the identity map and
therefore no other instantons of the type (\ref{eq:a2}) exist.

In the case of a SU(3) gauge group the $G_{i}$ (\ref{eq:18i})
are hermitian $3 \times 3$
matrices. The SU(3) embedding of the SU(2) instanton forms a
SU(3) instanton configuration
\be
G^{a}_{i}= \delta ^{a}_{i} \; , \hbo a=1,2,3 \hbo
G^{b}_{i}=0 \; , \hbo b=4,\ldots , 8 \; .
\label{eq:25i}
\en
This is the known SU(3) instanton and is widely used in
the instanton models describing the QCD vacuum.
This instanton corresponds to the $s=1/2$ representation of the
spin group and is illustrated in figure \ref{fig:2a}.
There also exists another non-trivial SU(3) instanton,
corresponding to the $s=1$ representation of the spin group, i.\ e.\ ,
%\footnote{ This fact was also realised by M.\ Schaden (private
%communication). }
\be
(\, G_{i} \, )_{kl} \; = \; -i \, \epsilon _{ikl}
\label{eq:26i}
\en
which is the maximal spin representation in $N=3$ dimensions.
These $G_{i}$ are up to a phase the antisymmetric generators
$t^{a}=2,5,7$ of the fundamental SU(3) representation, which also
forms the $\bar{3}$ representation of SU(3). For this  SU(3)
solution, the non-zero components read (see figure \ref{fig:2a})
\be
G^{7}_{1} \; = \; - \, G^{5}_{2} \; = \; G^{2}_{3} \; = \; 2 \; .
\label{eq:27i}
\en
This solution thus has a non-trivial SU(3) color structure and
from (\ref{eq:23i}) it carries winding number $n=4$.
Since in three dimensions (besides the trivial $s=0$) only the
$s=1/2$ and $s=1$ representations of SU(2) can be realised there are no
further self-dual SU(3) instantons of type (\ref{eq:a2}).

For a SU(4) gauge group the embeddings of the SU(3) instantons
discussed above are trivially SU(4) instanton solutions. A
SU(4) instanton with non-trivial SU(4) color structure again arises
from the identification of the $G_{i}$'s with the maximal spin
$s=3/2$ representation of the SU(2) spin group in $N=4$ dimensions.
According to (\ref{eq:29i}) this solution
carries winding number $n=10$. Two further SU(4) solutions are provided
by the two $s=1/2$ representations of SU(2) in four dimensions
given by  the  't~Hooft's symbols
\be
G_{k} \; = \; - \frac{i}{2} \eta ^{k}  \hbox to 2cm {\hfill or \hfill }
G_{k} \; = \; - \frac{i}{2} \bar{\eta }^{k}
\label{eq:28i}
\en
where the $\eta^{k}_{\mu \nu }$, $\bar{\eta }^{k}_{\mu \nu }$
are now matrices in color space.
These instantons correspond to the $(\frac{1}{2}, 0)$ and
$(0, \frac{1}{2})$ representations of the SO(4)$\sim $SU(2)$\times $SU(2)
subgroup of SU(4).
{}From (\ref{eq:21i}) and (\ref{eq:23i}) we see that these instantons carry
winding number $n=2$.
Of course, to any self-dual instanton solution with
winding number $n$ found above there exists an anti self-dual
instanton with winding number $-n$ constructed by exchanging
$G_{i}$ and $\bar{G}_{i}$.

\bigskip
{\it 4.\ Exotic Skyrmions }

\medskip
In the context of large-$N$ QCD~\cite{tho73}
baryons are described as topological solitons of the chiral meson
field. These solitons are commonly referred to as Skyrmions (see e.g.\
\cite{skyrme}). They represent mappings $U(\vec{x})$
from ${\cal R}^{3}$ into the flavour group SU(N) with $U(\vec{x})
\rightarrow 1 $ as $\mid \vec{x} \mid \rightarrow \infty$.
This compactification condition is required for finite energy
of the solitons and implies that $U(\vec{x})$ has a well defined degree
(winding number), which is identified~\cite{sky62,wi83} with
the baryon number
\be
B \; = \; \frac{1}{24 \pi ^{2} } \, \epsilon _{ijk} \, \tr
L_{i} L_{j} L_{k} \; , \hbo
L_{i} \; = \; U^{\dagger } \partial _{i} U \; .
\en
In the effective meson theories (e.g.\ the Skyrme model~\cite{skyrme}
or more microscopic models~\cite{njl}) the lowest energy baryon number
$B=1$ soliton configuration has the hedgehog form
\be
U(\vec{x}) \; = \; \exp \bigl( i \theta (r) \hat{x} \vec{\tau } \bigr) \; ,
\label{eq:hedge}
\en
where $\vec{\tau}$ denotes the Pauli isospin matrices or their
embeddings into SU($N>2$) flavour groups.

Atiyah and Manton have constructed SU(2) Skyrmion fields~\cite{manton}
by computing the holomony of SU(2) Yang-Mills instantons
$A_{\mu }(x)$ in ${\cal R}^{4}$ along the lines parallel to the
time-axis
\be
U(\vec{x}) \; = \; \hbox{P} \, \exp \bigl( i
\int _{ - \infty }^{ \infty } d x_{4} \; A_{4}(\vec{x}, x_{4} ) \bigr)
\; , \hbo A_{\mu }(x) \; = \; A^{a}_{\mu }(x) \tau ^{a} \; .
\label{eq:x}
\en
Here P denotes path ordering. This procedure can be straightforwardly
extended to the SU(N) instantons obtained in the present letter.
Inserting for $A_{4}$ the explicit form given by (\ref{eq:a3},
\ref{eq:14i}, \ref{eq:32}) the chiral field resulting from (\ref{eq:x})
becomes
\be
U(\vec{x}) \; = \; \exp \bigl( i \, 2 \theta (r) \vec{G} \vec{x} \bigr) \; ,
\hbo (\vec{G})_{k} \; = \; G^{a}_{k} t^{a} \; = \; G_{k} \; .
\label{eq:x1}
\en
It has a generalised hedgehog form with a profile function
\be
\theta (r) \; = \; \pi \, \bigl( 1 \, - \, \frac{1}{
\sqrt{ 1 + (\rho / r )^{2} } } \bigr) \; ,
\label{eq:pro}
\en
where $\rho $ is the (four dimensional) instanton radius.
For $N>2$ the flavour structure of the Skyrmion field (\ref{eq:x1})
is obviously different from the trivial SU(N) embedding of the
heghog (\ref{eq:hedge})
\footnote{ For $N=2$ the maximum spin instanton coincides with the
standard Polyakov-'t~Hooft instanton instanton and we recover
Atiyah and Manton's result. }. Nevertheless
the profile function (\ref{eq:pro}) is the same as found by Atiyah and
Manton~\cite{manton} in the SU(2) case. This is due to the fact that
the matrices $G_{k}$ form a SU(2) subalgebra (see (\ref{eq:34})) of the
SU(N) group.
The instanton radius $\rho $ can be tuned to minimise the classical
soliton energy. For the Skyrme model the SU(2) chiral field
constructed in this way from the Polyakov-'t~Hooft instanton
has energies exceeding the exact soliton energy by less than
1\% \cite{manton}.

In the Atiyah-Manton-procedure
the degree of the field $U(\vec{x})$ coincides with the
Pontryagin index of the generating instanton. Thus the maximal spin
instanton $s=1$ for SU(3) gives rise to a Skyrmion (\ref{eq:x1})
with baryon number $B=4$ living in SU(3) flavour space and thus
carrying strangeness, i.e.\ a strange $\alpha $-particle.
Finally let us consider the SU(4) instantons (\ref{eq:28i}) with $s=2$.
They give rise to a dibaryonic soliton which carries beside
strangeness also charm. It would be interesting to work out
physical properties of these exotic SU(n) Skyrmions.

\vspace{ 2 true cm }
{\bf Figure 1: }
The embedding of the Polyakov-t'Hooft instanton (a) in SU(3),
   and the maximum spin instanton (b) of SU(3).
\end{document}